\documentclass[aps,prb,superscriptaddress,twocolumn,groupedaddress, showpacs]{revtex4}

\usepackage{graphicx}
\usepackage{amsmath}
\usepackage{amsfonts}
\usepackage{amssymb}
\usepackage{units}
\usepackage{color}

\begin{document}

\title{Roughness-induced magnetic decoupling at organic-inorganic interface}

\author{Hiroki Ono}
\affiliation{Graduate School of Engineering and School of Engineering, Nagoya University, Furo-cho, Chikusa-ku, Nagoya, Aichi 464-8603, Japan}

\author{Yoshitaka Umeda}
\affiliation{Graduate School of Engineering and School of Engineering, Nagoya University, Furo-cho, Chikusa-ku, Nagoya, Aichi 464-8603, Japan}

\author{Kaito Yoshida}
\affiliation{Graduate School of Engineering and School of Engineering, Nagoya University, Furo-cho, Chikusa-ku, Nagoya, Aichi 464-8603, Japan}

\author{Kenzaburo Tsutsui}
\affiliation{Graduate School of Engineering and School of Engineering, Nagoya University, Furo-cho, Chikusa-ku, Nagoya, Aichi 464-8603, Japan}

\author{Kohei Yamamoto}
\affiliation{Department of Materials Molecular Science, Institute for Molecular Science, Okazaki 444-8585, Japan}

\author{Osamu Ishiyama}
\affiliation{Department of Materials Molecular Science, Institute for Molecular Science, Okazaki 444-8585, Japan}

\author{Hiroshi Iwayama}
\affiliation{Department of Materials Molecular Science, Institute for Molecular Science, Okazaki 444-8585, Japan}

\author{Eiken Nakamura}
\affiliation{Department of Materials Molecular Science, Institute for Molecular Science, Okazaki 444-8585, Japan}

\author{Toshihiko Yokoyama}
\affiliation{Department of Materials Molecular Science, Institute for Molecular Science, Okazaki 444-8585, Japan}

\author{Masaki Mizuguchi}
\affiliation{Graduate School of Engineering and School of Engineering, Nagoya University, Furo-cho, Chikusa-ku, Nagoya, Aichi 464-8603, Japan}
\affiliation{Institute of Materials and Systems for Sustainability (IMaSS), Nagoya University, Furo-cho, Chikusa-ku, Nagoya, Aichi 464-8601, Japan}

\author{Toshio Miyamachi}
\email[]{toshio.miyamachi@imass.nagoya-u.ac.jp}
\affiliation{Graduate School of Engineering and School of Engineering, Nagoya University, Furo-cho, Chikusa-ku, Nagoya, Aichi 464-8603, Japan}
\affiliation{Institute of Materials and Systems for Sustainability (IMaSS), Nagoya University, Furo-cho, Chikusa-ku, Nagoya, Aichi 464-8601, Japan}


\begin{abstract}
We have investigated structural, electronic and magnetic properties of H$_2$Pc on Fe$_2$N/Fe using low-energy electron diffraction and soft x-ray absorption spectroscopy/x-ray magnetic circular dichroism. Element specific magnetization curves reveal that the magnetic coupling with H$_2$Pc enhances the perpendicular magnetic anisotropy of Fe$_2$N/Fe at the H$_2$Pc coverage of 1 molecular layer. However, adding two and three molecular layers of H$_2$Pc reverts the shape of magnetization curve back to the initial state before H$_2$Pc deposition. We successfully link appearance and disappearance of the magnetic coupling at the H$_2$Pc-Fe$_2$N/Fe interface with the change of hybridization strength at N sites accompanied by the increase in the H$_2$Pc coverage.
\end{abstract}

\date{\today}
\maketitle

  Organic materials are expected to be useful to spintronics devices because of long spin lifetime derived from their intrinsic weak spin-orbit coupling. Especially, an organic-inorganic hybrid interface has attracted much attention due to its controllable interfacial spin states via electronic hybridization. \cite{Sanvito2010}. Recent spin-dependent electron transport measurements reveal the importance of not only electronic hybridization but also local structures at the organic-inorganic interface on the emergence of novel spin functionalities \cite{Nakayama2018, Isshiki2019}. In the case of organic molecules on a magnetic substrate, light elements such as H, C, N, O, which constitute an organic molecule, can magnetize via magnetic coupling with the substrate, playing an important role in characterizing magnetic properties of the whole system \cite{Djeghloul2013, Gueddida2016}. Since the hybridization strength considerably depends on molecular adsorption structure, precise control of the magnetic coupling at the single molecule level is required. However, an experimental study based on the hybridization of light elements in an organic molecule with magnetic substrate has been rarely done.

  Here we investigate the magnetic coupling of organic-inorganic interface consisting of metal-free phthalocyanine (H$_2$Pc) molecules and a bilayer $\gamma$'-Fe$_4$N (Fe$_2$N/Fe) on Cu(001) (see Fig.1 (a)). In contrast to imperfect layer-by-layer growth of pristine Fe atomic layers on Cu(001) \cite{Biedermann2004}, Fe$_2$N/Fe grows with atomically flat and homogeneous surface due to its high lateral lattice stability \cite{Takahashi2017, Hattori2018}. This allows to selectively extract the impact of molecular adsorption geometry on the magnetic coupling. H$_2$Pc is a prototypical planar molecule, and chosen as an organic molecule in this work since the efficient hybridization with a magnetic substrate is expected due to its preferential $\pi^*$-conjugated configuration \cite{Schmaus2011}. In addition, An H$_2$Pc molecule consists only of H, C, N, and thus suitable to investigate the magnetic coupling with Fe$_2$N/Fe in terms of light elements. The results demonstrate that the formation of the magnetic coupling between H$_2$Pc and Fe$_2$N/Fe can clearly appear as the increase in the hybridization strength at N sites.

  H$_2$Pc/Fe$_2$N/Fe hybrid thin films were grown on Cu (001) in ultrahigh vacuum (UHV) with the base pressure better than 2 $\times$ 10$^{-10}$ Torr. A cyclic process of Ar$^+$ ion sputtering and annealing at $\sim$ 700 K of Cu(001) was first conducted to prepare atomically flat and clean substate surface. Fe$_2$N/Fe was then fabricated by depositing Fe using an electron-bombardment-type evaporator onto N$^+$ ion bombarded Cu(001) with an energy of 0.5 keV, and by subsequent annealing at $\sim$ 600 K. Finally, H$_2$Pc molecules were deposited on Fe$_2$N/Fe/Cu(001) at 600 K in the base pressure better than 5$\times$10$^{-10}$ Torr, while keeping the substrate at room temperature. Surface structures, element-specific electronic and magnetic properties of H$_2$Pc/Fe$_2$N/Fe were investigated by low-energy electron diffraction (LEED) and soft x-ray absorption spectroscopy/x-ray magnetic circular dichroism (XAS/XMCD). XAS/XMCD measurements were performed at BL 4B of UVSOR-III in a total electron yield mode at 8 K and B = 0 - $\pm$ 5 T in the base pressure of $\sim$ 1 $\times$10$^{-10}$ Torr \cite{Nakagawa2008}. XAS spectra were recorded at $\theta$ between 0$^{\circ}$ and 75$^{\circ}$, which is the incident x-ray angle with respect to the surface normal. The XMCD spectrum was defined as $\mu_+$ – $\mu_-$ recorded at the normal (NI: $\theta$ = 0$^{\circ}$) and the grazing (GI: $\theta$ = 55$^{\circ}$) geometries. Note that $\mu_+$ and $\mu_-$ were the XAS spectra with the photon helicity parallel and antiparallel to the magnetic field direction, respectively. The circular polarization of the x-ray was set to $\sim$ 65 $\%$. The coverages of Fe$_2$N/Fe and H$_2$Pc were determined by the quartz-crystal microbalance (QCM) and XAS edge jump \cite{Takahashi2017, Takagi2010}.

  Structural properties of H$_2$Pc/Fe$_2$N/Fe hybrid thin films are characterized by XAS and LEED. Figure 1(b) displays XAS spectra of Fe and Cu L edges of Fe$_2$N/Fe on Cu(001). The Fe coverage can be evaluated with sub-monolayer accuracy from the ratio of the XAS edge jump of Fe to that of Cu (Fe$_L$/Cu$_L$). Considering Fe$_L$/Cu$_L$ of $\sim$ 0.1 as the Fe coverage of a single atomic layer \cite{Takahashi2017, Kawaguchi2020, Kawaguchi2022}, Fe$_L$/Cu$_L$ of $\sim$ 0.2 in this work ensures controlled Fe deposition by QCM to fabricate Fe$_2$N/Fe. We further confirm that the shape and relative intensity of L$_3$ and L$_2$ peaks of Fe XAS spectrum are almost unchanged after deposition of 1, 2 and 3 molecular layer (ML) of H$_2$Pc as shown in the inset of Fig. 1(b). This indicates that the structural transition from Fe$_2$N/Fe to Fe/Fe assisted by the nitrogen surfactant effect \cite{Takahashi2017, Kawaguchi2020, Kawaguchi2022} is absent and the Fe$_2$N lattice is preserved after H$_2$Pc deposition since magnetic moments derived from Fe 3d electronic states distributed near the Fermi energy (E$_F$) are quite different between Fe$_2$N/Fe and Fe/Fe on Cu(001) \cite{Takahashi2017, Abe2008}.
  
\begin{figure}[]
\begin{center}
\includegraphics[width=8cm]{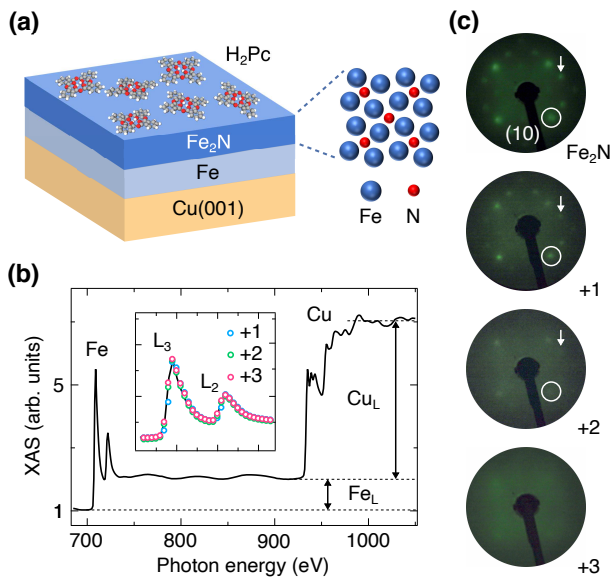} 
\caption{(a) Schematics of H$_2$Pc/Fe$_2$N/Fe hybrid thin films (left panel) and the p4gm(2$\times$2) reconstruction in the Fe$_2$N layer (right panel). Blue and red spheres represent Fe and N atoms, respectively. (b) Fe and Cu L$_{2,3}$ XAS spectra of Fe$_2$N/Fe on Cu(001) recorded in the NI geometry. The inset displays Fe L$_{2,3}$ XAS spectra of bare Fe$_2$N/Fe (solid line), and with 1 (blue), 2 (green) and 3 (pink) ML H$_2$Pc, respectively. (c) LEED patterns of bare Fe$_2$N/Fe on Cu(001), and with 1, 2 and 3 ML H$_2$Pc (from top to bottom). The incident energy is set to 130 eV. Circles and arrows represent  LEED spots derived from the p(1$\times$1) lattice and p4gm(2$\times$2) superlattice.}
\end{center}
\end{figure}

  Figure 1(c) displays a series of LEED patterns of Fe$_2$N/Fe on Cu(001) with H$_2$Pc overlayers (0, 1, 2 and 3 ML). The incident electron energy of 130 eV leads to the LEED probing depth of at most several atomic layers. The LEED pattern of bare Fe$_2$N/Fe shows the p4gm(2$\times$2) reconstructed structure (see right panel of Fig.1(a)), which is characteristic of the topmost Fe$_2$N layer as previously reported \cite{Takahashi2017, Takagi2010, Takahashi2016}. We find that the p4gm(2$\times$2) LEED pattern is observed after deposition of 1ML H$_2$Pc. The appearance of the p4gm(2$\times$2) LEED pattern is a clear evidence of the existence of the Fe$_2$N layer underneath 1 ML H$_2$Pc, and is in line with observed H$_2$Pc coverage dependence of the Fe XAS spectra as shown in the inset of Fig. 1(b). With increasing the coverage of H$_2$Pc to 2 ML, LEED spots derived from the p4gm(2$\times$2) superlattice in addition to the p(1$\times$1) lattice are still recognizable. At the coverage of 3 ML H$_2$Pc, no clear LEED spot is observed. It is important to note that no additional superstructure spots derived from H$_2$Pc are observed throughout LEED observations in the energy range from 14 to 150 eV. In addition, the LEED background intensity is significantly enhanced as the coverage of H$_2$Pc increases. The above results provide the information on the growth of H$_2$Pc/Fe$_2$N/Fe hybrid thin films; H$_2$Pc molecules stack on Fe$_2$N/Fe in a disordered manner and the roughness of surface and interface considerably increases with increasing the H$_2$Pc coverage.

\begin{figure}[]
\begin{center}
\includegraphics[width=7cm]{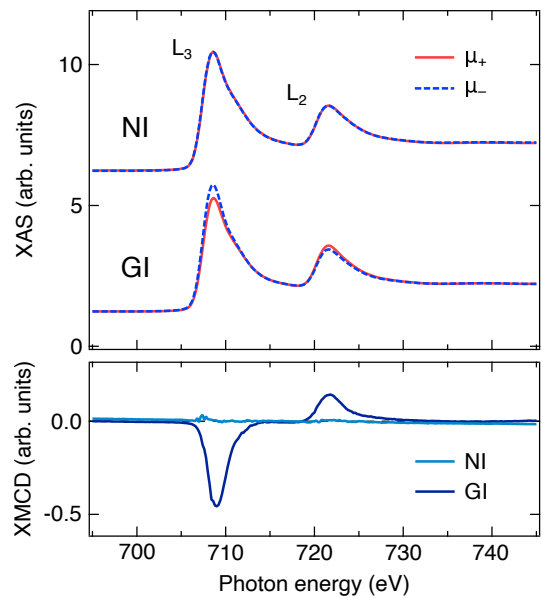} 
\caption{Fe L$_{2,3}$ remanent XAS and XMCD spectra of Fe$_2$N/Fe on Cu(001) recorded in the NI and GI geometries. After application of magnetic field up to $\pm$ 5 T, the XAS spectra are recorded at remanence conditions ($\pm$ 5 $\rightarrow$ 0 T). All the XAS spectra are normalized by the edge jump intensity.}
\end{center}
\end{figure}

  For XAS/XMCD measurements, we first check magnetic properties of bare Fe$_2$N/Fe on Cu(001). Figure 2 displays Fe L edge XAS ($\mu_+$, $\mu_-$) and XMCD ($\mu_+$ - $\mu_-$) spectra recorded at remanence conditions in the NI and GI geometries. While the remanent XMCD signal is observed in the GI geometry, that is almost negligible in the NI geometry, which clearly reveals the strong in-plane magnetic anisotropy of Fe$_2$N/Fe on Cu(001). In addition, the spin magnetic moment of Fe$_2$N/Fe is evaluated to be 1.3 $\pm$ 0.1 $\mu_B$/atom by XMCD sum rules \cite{Thole1992, Chen1995}. These results are in good agreement with the combined experimental and theoretical study of the same system \cite{Takahashi2017}, and thus ensure the high quality of Fe$_2$N/Fe in this work. Note that the average number of Fe 3d holes, n$_{hole}$, of 3.22 was used in the sum rule analysis \cite{Nakashima2019}. 

\begin{figure*}[]
\begin{center}
\includegraphics[width=12cm]{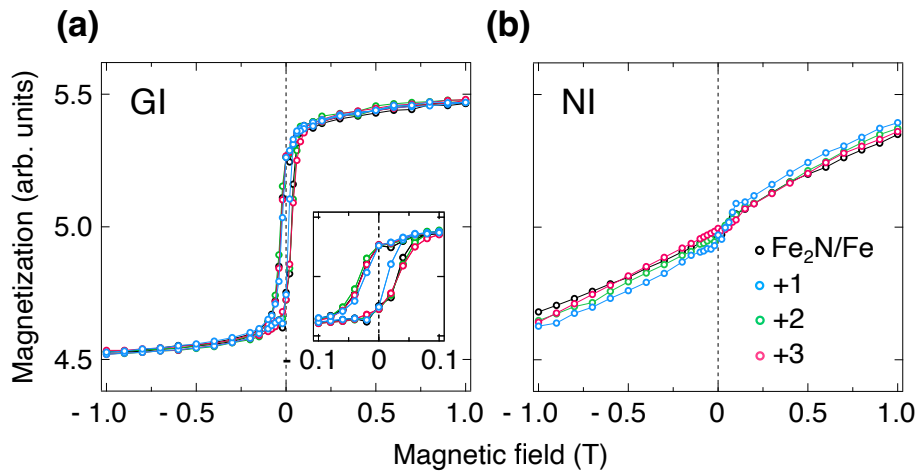} 
\caption{Magnetization curves of Fe$_2$N/Fe with 0 (black), 1 (blue), 2 (green) and 3 (pink) ML H$_2$Pc recorded in the (a) GI and (b) NI geometries, respectively. The L$_3$ peak intensity in the Fe XAS spectrum normalized by the L$_2$ one is plotted as a function of magnetic field.}
\end{center}
\end{figure*}  

  To investigate details of the magnetic coupling at the interface of H$_2$Pc and Fe$_2$N/Fe, we element-specifically measure Fe magnetization curves. Figure 3(a) displays magnetization curves of Fe$_2$N/Fe with 0, 1, 2, and 3 ML H$_2$Pc recorded in the GI geometry. A hysteresis loop with saturation magnetization of $\sim$ $\pm$ 1T is observed for bare Fe$_2$N/Fe, reflecting its in-plane magnetic easy axis. We find that the saturation magnetization of Fe$_2$N/Fe, which is dominantly attributed to the amplitude of the spin magnetic moment, is almost unchanged after adding H$_2$Pc overlayers. Indeed, the evaluated spin magnetic moment of Fe$_2$N/Fe with 1, 2, and 3 ML H$_2$Pc is 1.2 $\pm$ 0.1 $\mu_B$/atom, which is close to that of bare Fe$_2$N/Fe (1.3 $\pm$ 0.1 $\mu_B$/atom). The impacts of H$_2$Pc overlayers on the Fe magnetization curve can be seen in the coercivity near zero magnetic field. As shown in the inset of Fig. 3(a), a slight decrease in the coercivity is observed for Fe$_2$N/Fe with 1 ML H$_2$Pc. However, interestingly, adding 2 ML H$_2$Pc nearly recovers the coercivity of Fe$_2$N/Fe to its initial condition, i.e., bare Fe$_2$N/Fe, and no further changes are recognizable at the H$_2$Pc coverage of 3 ML. Since the coercivity is determined by various factors such as magnetic anisotropy, formation of magnetic domains, and domain wall pinning \cite{Hubert1998}, it is not straightforward to explain observed H$_2$Pc coverage dependence only from the Fe magnetization curves recorded in the GI geometry. Therefore, we further investigate its origin from Fe magnetization curves recorded in the NI geometry. 

  Figure 3(b) displays magnetization curves of Fe$_2$N/Fe with 0, 1, 2, and 3 ML H$_2$Pc recorded in the NI geometry. Due to the strong in-plane magnetic anisotropy of Fe$_2$N/Fe, all the Fe magnetization curves exhibit no remanence. As for the H$_2$Pc coverage dependence, the Fe magnetization curves show a tendency similar to those recorded in the GI geometry; the shape of the Fe magnetization curve changes only when 1ML of H$_2$Pc is deposited on Fe$_2$N/Fe. Concretely, the Fe magnetization curve becomes steeper with 1ML H$_2$Pc. This indicates that adding 1ML H$_2$Pc enhances the perpendicular magnetic anisotropy of Fe$_2$N/Fe, which rationalizes decreased coercivity observed in the GI geometry. Since all H$_2$Pc molecules are expected to be directly anchored to Fe$_2$N/Fe at the H$_2$Pc coverage of 1 ML even though they might be randomly adsorbed on the surface (see Fig.1 (c)), it is reasonable to ascribe the enhancement of the perpendicular magnetic anisotropy of Fe$_2$N/Fe to the magnetic coupling with H$_2$Pc. Further changes in the Fe magnetization curve by stacking 2 and 3 ML H$_2$Pc can be explained in terms of the relationship between the magnetic coupling and local structure at the interface of H$_2$Pc and Fe$_2$N/Fe. LEED observations shown in Fig.1 (c) reveal that the surface/interface roughness in this system significantly increases with increasing the H$_2$Pc coverage. We therefore presume that the second (and third) H$_2$Pc layer disrupts the molecule–surface bonding of H$_2$Pc molecules directly anchored to Fe$_2$N/Fe and inclines them via intermolecular interactions. This could lead to magnetic decoupling of the first H$_2$Pc layer from Fe$_2$N/Fe. Note that similar effects are reported for spin transport measurements of PbPc/Cu/NiFe system \cite{Isshiki2019}. However, the hybridization between PbPc and Cu inducing a Rashba spin splitting at the interface is relatively robust against the increase in the PbPc coverage, which is in contrast to drastic changes of the degree of the magnetic coupling between H$_2$Pc and Fe$_2$N/Fe. This indicates that the degree of the magnetic coupling of organic molecules and inorganic materials with localized d states near the E$_F$, i.e., Fe 3d states in this work, is quite sensitive to changes of local structure.

  To validate the scenario mentioned above, we focus on the H$_2$Pc coverage dependence of N K‐edge XAS. Figure 4(a) displays N K‐edge XAS spectra recorded at $\theta$ = 65$^{\circ}$ for 1, 2, 3 ML H$_2$Pc on Fe$_2$N/Fe, and bare Fe$_2$N/Fe as a reference. As seen for H$_2$Pc thin films on various substrates \cite{Annese2008, Willey2013}, the spectrum consists of two characteristic energy regions; $\pi^*$ resonances below $\sim$ 404 eV with peaks derived from unoccupied molecular orbitals, LUMO (A), LUMO+1 (B) and LUMO+2 (C) at 398, 400, and 402 eV, and $\sigma^*$ resonances with broad features at higher energies. Note that the pronounced $\pi^*$ resonances observed at $\theta$ = 65$^{\circ}$, where the polarization vector of x-ray is approximately perpendicular to the surface in this study, considerably decrease at $\theta$ = 15$^{\circ}$ (see SI). This indicate that, in spite of random and disorder stacking as revealed by LEED observations, H$_2$Pc molecules more or less maintain planar $\pi^*$-conjugated configurations with Fe$_2$N/Fe. In advance to discuss how the magnetic coupling modifies electronic structures of H$_2$Pc molecules at N sites, we subtract the N K‐edge XAS spectrum of bare Fe$_2$N/Fe as a background from the one of 1, 2, 3 ML H$_2$Pc on Fe$_2$N/Fe (see SI), which allows an accurate comparison among them. Figure 4(b) displays $\pi^*$ resonances in the background-subtracted N K‐edge XAS spectra of 1, 2, 3 ML H$_2$Pc on Fe$_2$N/Fe. We find that especially peaks B and C are broader for 1 ML H$_2$Pc, but become sharper for 2 ML H$_2$Pc. As for 3 ML H$_2$Pc, the spectral shape is almost identical to that of 2 ML H$_2$Pc. The magnetic coupling arises as a consequence of electronic hybridization between H$_2$Pc and Fe$_2$N/Fe. The hybridization of molecular orbitals with substrate can delocalize LUMO \cite{Miyamachi2012}, resulting in the broadening of peaks in the N K‐edge XAS spectrum \cite{Annese2008}. Thus, broad peaks observed for 1 ML H$_2$Pc is a clear signature of the efficient hybridization, i.e., magnetic coupling with Fe$_2$N/Fe. Furthermore, narrowing of the peak width with increasing the H$_2$Pc coverage to 2, and 3 ML is involved in weakening of the hybridization with Fe$_2$N/Fe, which is consistent with the magnetic decoupling at the interface. These results fully support the conclusions from LEED and XMCD observations. Note that the enhancement of the perpendicular magnetic anisotropy of Fe$_2$N/Fe with 1 ML H$_2$Pc is reasonable also in terms of density of states of Fe$_2$N/Fe and H$_2$Pc. Among 3d states of Fe$_2$N on Fe, out-f-plane oriented orbitals, especially d$_{3z^2 - r^2}$ and d$_{yz}$ are prominent at the energy range of about 0-3 eV above E$_F$ \cite{Takahashi2017}, where LUMO (A), LUMO+1 (B) and LUMO+2 (C) of H$_2$Pc exist. This leads to the broadening of the bandwidth of the out-f-plane 3d orbitals via the hybridization at the interface, which in turn can enhance the out-of-plane magnetization of Fe$_2$N/Fe triggered by spin-orbit interaction \cite{Stohr1999, Bruno1989}.

\begin{figure}[]
\begin{center}
\includegraphics[width=7cm]{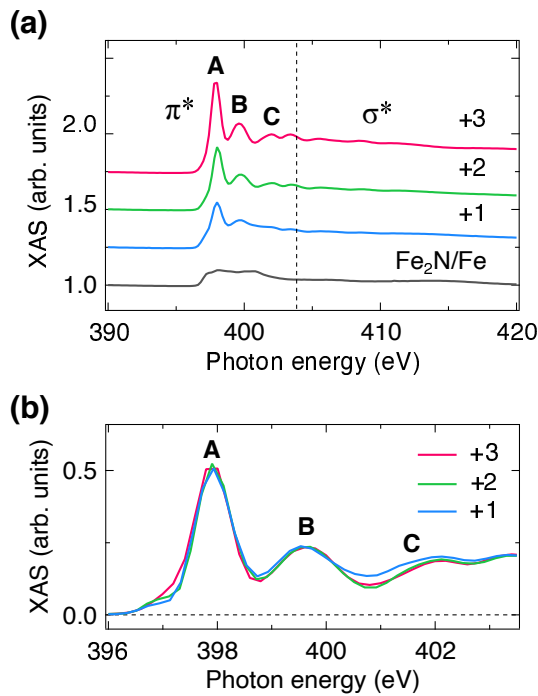} 
\caption{(a) N K‐edge XAS spectra of 0, 1, 2, 3 ML H$_2$Pc on Fe$_2$N/Fe (from bottom to top). (b) N K‐edge XAS spectra of 1, 2, 3 ML H$_2$Pc after subtracting Fe$_2$N/Fe background spectrum. The spectra are normalized by peak intensities and edge jump.}
\end{center}
\end{figure}

  In conclusion, we have investigated the H$_2$Pc coverage dependence of magnetic properties of Fe$_2$N/Fe on Cu(001). The results have revealed that the magnetic coupling enhances the perpendicular magnetic anisotropy of Fe$_2$N at the H$_2$Pc coverage of 1 ML, but adding 2 and 3 ML H$_2$Pc considerably increases surface/interface roughness of the system and accordingly decouples the magnetic interaction. We have additionally demonstrated the importance of light elements in a organic molecule to elucidate the correlation between structural and electronic/magnetic properties on the level of a single molecule toward utilizing the organic-inorganic interface in molecular spintronics devices.

\section*{Acknowledgment}
  This work was partly supported by JSPS KAKENHI Grant Numbers, JP20H05179, 20K21119, JP21H05016, 22H01957 and JP23H01849. A part of this work was conducted at the BL4B of UVSOR Synchrotron Facility, Institute for Molecular Science supported by Nanotechnology Platform Program $<$Molecule and Material Synthesis$>$ (JPMXP09S19MS2001b, JPMXP09S19MS2001, JPMXP09S20MS2001, JPMXP09S21MS2001, JPMXP09S21MS2002), and by Advanced Research Infrastructure for Materials and Nanotechnology in Japan (JPMXP1222MS2001, JPMXP1222MS2003) of the Ministry of Education, Culture, Sport, Science and Technology (MEXT), Japan.




\begin{thebibliography}{9}

\bibitem{Sanvito2010}S. Sanvito, Nature Phys {\bf47}, 562–564 (2010). 

\bibitem{Nakayama2018}H. Nakayama, T. Yamamoto, H. An, K. Tsuda, Y. Einaga, K. Ando, Sci. Adv. {\bf4}, eaar3899 (2018). 

\bibitem{Isshiki2019}H. Isshiki, K. Kondou, S. Takizawa, K. Shimose, T. Kawabe, E. Minamitani, N. Yamaguchi, F. Ishii, A. Shiotari, Y. Sugimoto, S. Miwa, Y. Otani, Nano. Lett. {\bf19}, 7119-7123 (2019).

\bibitem{Djeghloul2013}F. Djeghloul, F. Ibrahim, M. Cantoni, M. Bowen, L. Joly, S. Boukari, P. Ohresser, F. Bertran,
P. Le F\'{e}vre, P. Thakur, F. Scheurer, T. Miyamachi, R. Mattana, P. Seneor, A. Jaafar, C. Rinaldi, S. Javaid, J. Arabski, J. -P Kappler, W. Wulfhekel, N. B. Brookes, R. Bertacco, A. Taleb-Ibrahimi, M. Alouani, E. Beaurepaire, W. Weber, Sci. Rep. {\bf3}, 1272 (2013). 

\bibitem{Gueddida2016}S. Gueddida, M. Gruber, T. Miyamachi, E. Beaurepaire, W. Wulfhekel, M. Alouani, J. Phys. Chem. Lett {\bf7}, 900-904 (2016). 

\bibitem{Biedermann2004}A. Biedermann, R. Tscheliessnig, M. Schmid, P. Varga, Appl. Phys. A {\bf 78}, 807 (2004). 

\bibitem{Takahashi2017}Y. Takahashi, T. Miyamachi, S. Nakashima, N. Kawamura, Y. Takagi, M. Uozumi, V. N. Antonov, T. Yokoyama, A. Ernst, F. Komori, Phys. Rev. B {\bf95}, 224417 (2017).

\bibitem{Hattori2018}T. Hattori, T. Iimori, T. Miyamachi, F. Komori, Phys. Rev. Materials {\bf2}, 044003 (2018).

\bibitem{Schmaus2011}S. Schmaus, A. Bagrets, Y. Nahas, T. K. Yamada, A. Bork, M. Bowen, E. Beaurepaire, F. Evers, W. Wulfhekel, Nat. Nanotechnol. {\bf6}, 185-189 (2011).

\bibitem{Nakagawa2008}T. Nakagawa, Y. Takagi, Y. Matsumoto, T. Yokoyama, Jpn. J. Appl. Phys. {\bf47}, 2132 (2008).

\bibitem{Takagi2010}Y. Takagi, K. Isami, I. Yamamoto, T. Nakagawa, T. Yokoyama, Phys. Rev. B {\bf81}, 035422 (2010).

\bibitem{Kawaguchi2020}K. Kawaguchi, T. Miyamachi, T. Iimori, Y. Takahashi, T. Hattori, T. Yokoyama, M. Kotsugi, F. Komori, Phys. Rev. Materials {\bf4},1054403 (2020).

\bibitem{Kawaguchi2022}K. Kawaguchi, T. Miyamachi, T. Gozlinski, T. Iimori, Y. Takahashi, T. Hattori, K. Yamamoto, T. Koitaya, H. Iwayama, O. Ishiyama, E. Nakamura, M. Kotsugi, W. Wulfhekel, T. Yokoyama, F. Komori, Jpn. J. Appl. Phys {\bf61},SL1001 (2022).

\bibitem{Abe2008}H. Abe, K. Amemiya, D. Matsumura, J. Miyawaki, E. O. Sako, T. Otsuki, E. Sakai, T. Ohta, Phys. Rev. B {\bf77}, 054409 (2008).

\bibitem{Takahashi2016}Y. Takahashi, T. Miyamachi, K. Ienaga, N. Kawamura, A. Ernst, F. Komori, Phys. Rev. Lett. {\bf116}, 056802 (2016).

\bibitem{Thole1992}B. T. Thole, P. Carra, F. Sette, G. van der Laan, Phys. Rev. Lett. {\bf68}, 1943 (1992).

\bibitem{Chen1995}C. T. Chen, Y. U. Idzerda, H.-J. Lin, N. V. Smith, G. Meigs, E. Chaban, G. H. Ho, E. Pellegrin, F. Sette, Phys. Rev. Lett. {\bf75}, 152 (1995).

\bibitem{Nakashima2019}S. Nakashima, T. Miyamachi, Y. Tatetsu, Y. Takahashi, Y. Takagi, Y. Gohda, T. Yokoyama, F. Komori, Adv. Funct. Mater. {\bf29}, 1804594  (2019) . 

\bibitem{Hubert1998}A. Hubert and R. Sch\"{a}fer, {\it Magnetic domains} (Springer, Berlin, 1998).

\bibitem{Annese2008}E. Annese, J. Fujii, I. Vobornik, G. Panaccione, G. Rossi, Phys. Rev. B {\bf84}, 174443 (2011).

\bibitem{Willey2013}T. M. Willey, M. Bagge-Hansen, J. R. I. Lee, R. Call, L. Landt, T. van Buuren, C. Colesniuc, C. Monton, I. Valmianski, Ivan K. Schuller, J. Chem. Phys. {\bf139}, 034701 (2013).

\bibitem{Miyamachi2012}T. Miyamachi, M. Gruber, V. Davesne, M. Bowen, S. Boukari, L. Joly, F. Scheurer, G. Rogez,  T.K. Yamada, P. Ohresser, E. Beaurepaire, W. Wulfhekel, Nat. Commun. {\bf3}, 938 (2012).  

\bibitem{Stohr1999}J. St\"{o}hr, J. Magn. Magn. Mater {\bf200}, 470-497 (1999).

\bibitem{Bruno1989}P. Bruno, Phys. Rev. B {\bf39}, 865 (1989).

\end{thebibliography}
\end{document}